\documentclass[12pt]{JHEP3}

\usepackage{epsfig}

\newcommand{\half}{\frac12}

\newcommand{\del}{\partial}

\newcommand{\tbar}{{\bar t}}
\newcommand{\cO}{{\cal O}}
\newcommand{\cZ}{{\cal Z}}
\newcommand{\cF}{{\cal F}}
\newcommand{\cN}{{\cal N}}
\newcommand{\cT}{{\cal T}}
\newcommand{\cM}{{\cal M}}
\newcommand{\hX}{{\hat X}}

\def\be{\begin{equation}}
\def\ee{\end{equation}}
\def\bea{\begin{eqnarray}}
\def\eea{\end{eqnarray}}

\def\tr{{\bf\ tr\,}}

\def\ts{\textstyle}

\title{Topological Matrix Models, Liouville Matrix Model 
and $c=1$  String Theory\footnote{\em Based on lectures delivered 
at the IPM Workshop on String Theory, Bandar-e-Anzali, Iran, in October
2003.} }
\author{Sunil Mukhi\footnote{Email: \tt mukhi@tifr.res.in}\\
\it Tata Institute of Fundamental Research,\\
\it Homi Bhabha Rd, Mumbai 400 005, India}
\abstract{This is a review of some beautiful matrix models
related to the moduli space of Riemann surfaces as well as to
noncritical $c=1$ string theory at self-dual radius.  These include
the Penner model and the $W_\infty$ model, which have different
origins but are equivalent to each other. In the final section, which
is new material, it is shown that these models are also equivalent to
a Liouville matrix model. We speculate that this might be interpreted
in terms of $N$ D-instantons of the $c=1$ string.}
\preprint{hep-th/0310287\\
TIFR/TH/03-21}


\begin{document}

\section{Introduction}

In recent months there has been a revival of interest in the
noncritical $ c=1$ bosonic string\cite{McgVer,KleMalSei,McgTesVer}
(and its worldsheet supersymmetric
counterparts\cite{TakTou,DouKleKutMalMarSei,McgMurVer}). New insight
has been gained into this relatively simple string theory, using
several developments that came after the previous matrix revolution:
D-branes\cite{Polchinski}, M(atrix) Theory\cite{BFSS,IKKT},
AdS/CFT\cite{Maldacena} and the understanding of boundary states in
Liouville theory\cite{FatZamZam,Teschner,ZamZam}.

Most of the recent work has centred on matrix quantum mechanics, whose
double-scaled limit is believed to represent the $ c=1$ string (at
least perturbatively). The basic idea is that the matrix of this model
is the matrix-valued tachyon on the worldline of a collection of $N$
D0-branes.

There are also claims that a nonperturbatively consistent version of
matrix quantum mechanics can be formulated. This is supposed to be
equivalent not to the bosonic $ c=1$ string, but to a worldsheet
supersymmetric version, the ${\hat c}=1$ type 0B 
string\cite{TakTou,DouKleKutMalMarSei}. We will not
discuss this latter theory here. But many of the observations in the
present review presumably can, and should, be generalised to the
noncritical type 0B (and 0A) string background.

In its simplest (un-orbifolded) form, the $c=1$ string has a
translationally invariant space/time direction $X$. If it is
Euclidean, it can be compactified on a circle of radius $ R$.  In that
case, all physical quantities (partition function and amplitudes)
depend on $ R$. The Euclidean theory can be interpreted as a
finite-temperature theory with $R$ labelling the inverse temperature.

The value $ R=1$ (in units where $\alpha'=1$) is special because the $
c=1$ CFT is then self-dual under \be R\to {1\over R} \ee In this
article we will focus on this background: the Euclidean $ c=1$ string
with the $X$ direction compactified at the self-dual radius. For
short, we will refer to the theory as ``$c=1,R=1$''.

There are many indications that string theory is {\em topological} in
this background. The term ``topological string theory'' is usually
taken to mean a string theory where the matter sector has a twisted
$\cN=2$ superconformal algebra with central charge zero. To make it a
string theory, this matter then has to be coupled to topological
worldsheet gravity\cite{LabPerWit,Witten}, which can be described by a
set of bosonic and fermionic ghosts with total central charge
zero. The fermionic charge of the twisted $\cN=2$ algebra is the BRST
charge defining physical states.

A classic example of such a theory is a superstring on a Calabi-Yau
background, whose superconformal worldsheet theory has $\cN=2$
supersymmetry with central charge $c=9$.  On making
the standard topological twist of the superconformal algebra:
\be
T(z) \to T(z) + \half \del J(z)
\ee
where $J(z)$ is the $U(1)$ current, the algebra acquires central
charge 0 and we have a suitable matter system for a topological
string\cite{WittenMirror}. 

It is known that ordinary string theory, including the ghost sector,
can be considered as topological matter\cite{GatSem,BerLerNemWar,MukVaf},
which makes the distinction between topological and non-topological
theories less clear-cut. Even critical bosonic and superstrings are
known to be ``topological'' in this sense. The best distinction one
can make is that the latter theories are ``already twisted'' while the
conventional topological theories are formulated as $\cN=2$
superconformal theories and then given a topological twist.

Topological string theories are generally related to the topology of
the moduli space of Riemann surfaces. For example, ``pure''
topological gravity describes intersection theory on cohomology
classes associated to moduli space, or to vector bundles on moduli
space\cite{Witten}.

Some of the indications that $c=1, R=1$ is topological arise from its
relation to other theories. The partition function of $c=1,R=1$ is
closely related to that of the Penner matrix
model\cite{Penner,DisVaf}, a model constructed to count the Euler
characteristic of the moduli space of punctured Riemann surfaces.
Amplitudes in $c=1,R=1$ are summarised in the form of $W_\infty$
constraints\cite{DijMooPle} and the partition function of the
perturbed theory is a $\tau$-function of an integrable
hierarchy\footnote{A different approach to $W_\infty$ constraints and
associated matrix models is described for example in
Ref.\cite{BonXio}.}. This $\tau$-function in turn can be written as a
matrix model, the $W_\infty$ matrix model\cite{ImbMuk} (this has been
previously referred to as the ``Kontsevich-Penner
model''\cite{DijMooPle} and a ``Kontsevich-type model for
$c=1$''\cite{ImbMuk}).

Another indication of the topological nature of $c=1,R=1$ is that it
is dual to the topological 2d black
hole\cite{MukVaf,OogVaf,HorKap}. The latter theory indeed starts life
as an $\cN=2$ superconformal field theory in two dimensions, where the
second supersymmetry is a consequence of the Kazama-Suzuki
construction. This theory is also dual to topological Landau-Ginsburg
theory\cite{GhoMuk,HanOzPle,GhoImbMuk}, which is a convenient
formulation for explicitly computing amplitudes and comparing them to
those of the original $c=1,R=1$ theory as a test of the duality.

Finally, $c=1,R=1$ has been shown to be dual to topological strings on
a conifold singularity\cite{GhoVaf}. Recently it has also been
argued\cite{DijVaf} that $c=1,R=\infty$, which might be more
``physical'', is dual to an infinite-order orbifold of the
conifold. Such infinite-order orbifolds can be understood in the
language of deconstruction\cite{ArkCohGeo,HilPokWan}. In particular,
the above orbifold is believed to be dual to the (2,0) CFT in 6
dimensions\cite{ArkCohKapKarMot} (an alternative limit instead gives
rise to the nonrelativistic type IIA string\cite{MukRanVer}). Thus it
may be that the noncompact $c=1$ string can also be usefully
formulated as a topological theory. In this case one could even try to
continue the $X$ direction back to Minkowski signature and see what
happens in the dual theory. Progress on these issues is, however,
quite limited to date.

The various developments described above are important because the
``topological'' context is usually simpler than the ``dynamical''
one. Moreover, the former is often embedded in the latter, an example
being Gopakumar-Vafa duality\cite{GopVaf}, in which an open-closed
topological string duality can be lifted\cite{Vafa} to an open-closed
string duality in the full superstring theory. The computations in the
topological theory are related to specific types of amplitudes in the
dynamical theory.

In this article we will focus on the Penner and $W_\infty$
matrix models. These have been dubbed ``topological matrix models''
because they originate in the description of the moduli space of
Riemann surfaces and also because they are dual to $c=1,R=1$. These
are models of {\em constant matrices} and are quite different, in
spirit as well as in details, from the familiar $ c=1$ matrix quantum
mechanics. In particular, the nature and role of the large-$N$ limit
is rather different.

The plan of this article is as follows. In Section \ref{conerone} we
start by briefly reviewing the Liouville and the matrix quantum
mechanics approaches to $c=1$ string theory. After that  we collect some
explicit results from matrix quantum mechanics that will be useful in
the subsequent discussions.

In Section \ref{riemannpenner} we will explain how the Penner model is
related to the triangulation of the moduli space of Riemann surfaces,
rather than the triangulation of random surfaces. This requires some
discussion of meromorphic quadratic differentials on Riemann
surfaces. Section \ref{winf} deals with the $W_\infty$ model that is
built from correlators of $c=1,R=1$. The partition function is a
$\tau$-function of the Toda hierarchy. Relationships between this
model and the Penner model as well as the Kontsevich model are
discussed. Essentially the $W_\infty$ model is a perturbation of the
Penner model by tachyon couplings, but we will see that the relation
between them involves a somewhat subtle change of variables that may
have a physical significance.

To date, both these models have been constructed {\em a posteriori} by
computing appropriate amplitudes of the $c=1,R=1$ string from matrix
quantum mechanics, and then summing them up into an appropriate
generating function. This is a somewhat ad hoc state of affairs and
does not provide a fundamental explanation of how they arise, nor does
it indicate how these models can be generalised to other backgrounds.

In Section \ref{lastsec}, we will present some observations which might
help to remedy this situation. It will be shown that the $W_\infty$
model can be rewritten as a ``Liouville matrix model'' -- a model of
constant matrices with an exponential plus linear potential.
This way of rewriting the model is possible only because of its
specific form, including the dependence on coupling constants.

The presence of a matrix with an exponential plus linear potential is
strongly suggestive of D-instantons moving in the background of a
linear dilaton as well a cosmological (exponential) term arising from
the closed string tachyon. We will try to develop the analogy, but will
not give a conclusive argument that the $W_\infty$ model is
really the world-point action on $N$ D-instantons in the $c=1,R=1$
string theory. If this connection can really be demonstrated, it would
indicate a new type of holographic relationship, analogous to the IKKT
matrix model\cite{IKKT} of critical string theory. It might also point
to some new relationships between the perturbation series of string theory
and the topology of moduli space.

An important recent development is the construction of a new matrix
model called the ``normal matrix model'' (NMM)\cite{AleKazKos}. Like
the $W_\infty$ model, this one too is built directly out of the
$W_\infty$ solution for the partition function of $c=1,R=1$ (and also
other integer $R$). However, it does not seem to be known at present
what is the precise relation of NMM to the $W_\infty$ and Penner
models. We will briefly discuss the NMM at the end in order to exhibit
some similarities to the other models that are the main subject of
this article. Interestingly, in Ref.\cite{AleKazKos} a duality has
been proposed between matrix quantum mechanics and NMM, where the two
models are associated respectively to the non-compact and compact
cycles of a certain complex curve. This suggests an elegant answer to
the question of how $c=1$ matrix quantum mechanics and topological
matrix models are related. The discussion of the normal matrix model
presented here will unfortunately be very brief.

\section{$c=1$ at $R=1$}\label{conerone}

In this section we will summarise some results from the  matrix quantum
mechanics  approach to $ c=1$ string theory. An excellent review is
Ref.\cite{KlebanovReview}, where derivations of the formulae to be
presented below, as well as extensive references, can be found. First
we will briefly discuss the definition of $c=1$ string theory, both
from the continuum perspective and through matrices.

\subsection{Continuum formulation of $c=1$}

The continuum description of noncritical $c=1$ strings starts from a
matter conformal field theory of a free scalar field $X$, with
energy-momentum tensor:
\be
T^X_{zz} = - \del X\,\del X
\ee
This system has unit central charge, and its basic conformal fields
are the vertex operators
\be
V =~~ :e^{ikX}:
\ee
as well as polynomials in $\del^n X$ for various $n$, and products of
these with the vertex operators. The coordinate $X$ is taken to be the
time direction, though one also frequently studies a Euclideanised
theory where $X$ is taken to be spacelike.

A string theory with the critical central charge arises by coupling
the above CFT to a Liouville field with energy-momentum tensor:
\be
T^\phi = -\del \phi\,\del \phi + 2\, \del^2 \phi
\ee
This Liouville field has central charge $c=25$, so that the string is
in this sense critical. However, it is supposed that this field arises
as the scale factor of the  worldsheet metric, which adjusts itself
self-consistently to carry this central charge. The Liouville field is
spacelike.

Observables of this theory are obtained by computing the BRST
cohomology, just as is done for critical string theory. An important
class of observables are the ``tachyons'', which in Euclidean
signature are given by:
\be
T_{ k} = e^{(2-|k|)\phi}\, e^{i kX}
\ee
The name ``tachyon'' is a misnomer, as these are actually massless
states in two dimensions. 

There are also other physical modes in the BRST cohomology, the
so-called ``discrete states''\cite{Polyakov,LiaZuk,MukMukSen}, 
which arise at
special values of the momentum. These can be thought of as the two
dimensional analogues of the graviton and other tensor modes of a
closed string theory. In two dimensions, tensor fields have no
propagating field-theoretic degrees of freedom, but in this theory
they do have residual discrete modes that survive in the BRST
cohomology. The tachyons described above can scatter into these
discrete states, a fact which will become important when we discuss
amplitudes.

\subsection{Matrix formulation of $c=1$}

A very powerful description of $c=1$ strings can be obtained starting
from a random matrix integral:
\be
\int [dM]\,e^{{\ts-\beta\int_0^{2\pi R} 
dx\tr\left(\half \dot{M}^2 + {1\over
2}M^2 - 
{1\over 3!}M^3\right)}}
\ee
where the matrix measure is given by:
\be
[dM] \equiv \prod_i dM_{ii}\prod_{i<j}dM_{ij}\,dM^*_{ij}
\ee
A perturbative expansion of this integral leads to 't Hooft-type
Feynman diagrams with cubic vertices.  Each such diagram specifies a
unique Riemann surface topology on which it can be drawn as a random
lattice with three lines meeting at every point. The faces are
arbitrary $n$-gons. The dual lattice has $n$ lines meeting at a point
but the faces are triangles. The result is a triangulated Riemann
surface.
\smallskip

\EPSFIGURE{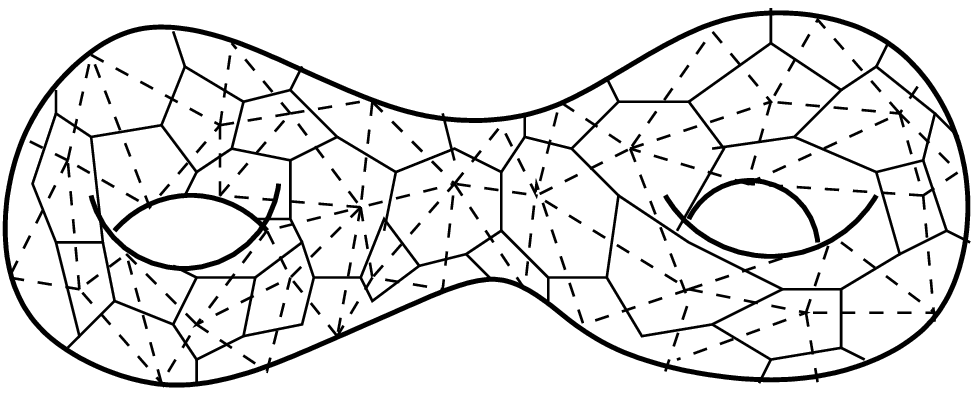}{A Riemann surface, with solid 
lines representing the matrix model diagram and dotted 
lines representing the dual triangulation.}

Thus the matrix quantum mechanics can be taken to represent the
discretisation of a Riemann surface. Summing over all triangulations
of the Riemann surface amounts to summing over all inequivalent
conformal classes, so the result should be equivalent to a string theory.
Indeed, this is the $c=1$ string. The time direction that we called
$X$ in the continuum case is associated to the time coordinate $t$ on
which the matrix coordinate depends, while the Liouville direction in
the continuum language is related to the infinite number of
eigenvalues of the matrix at large $N$\cite{DasJev}.

To solve this string theory, the matrix is first diagonalised. Then
the eigenvalue-dependent part of the measure becomes:
\be
[dM] = \prod_i d\lambda_i\, \Delta^2(\lambda)
\ee
where $\Delta(\lambda) = \prod_{i<j}(\lambda_i - \lambda_j)$ is the
Vandermonde determinant. In terms of eigenvalues, the kinetic term of
the Hamiltonian is:
\be
{1\over\Delta(\lambda)}\sum_i {d^2\over d\lambda_i^2}\Delta(\lambda_i)
\ee
acting on wave functions $\chi(\lambda_i)$. By redefining the wave
function as follows:
\be
\chi(\lambda_i) \to \Psi(\lambda_i)= \Delta(\lambda_i)\chi(\lambda_i)
\ee
we have a simpler Hamiltonian but the new wave function $\Psi$ is {\em
fermionic}, since interchange of any two eigenvalues gives a minus
sign. So the matrix eigenvalues behave as $N$ fermions moving in the
given potential.

The ground state of the theory is found by filling up the
first $N$ levels (after regulating the potential) to the Fermi level $
-\mu_F$. Next we take a double-scaling limit
\be
\mu_F\to 0,\quad \beta\to\infty
\ee
with $\mu=\beta\mu_F$ fixed.
In this limit, for purposes of perturbation theory the eigenvalues
behave  like fermions in an inverted harmonic oscillator potential.

The physical quantities (partition function, correlation
functions) of this model will be given  perturbatively  as an
expansion in powers of 
\be 
g_s^2\equiv {1\over \mu^2}
\ee
 Nonperturbative  effects will typically arise as terms like 
\be
e^{-\ts 1/g_s}\sim e^{-\ts \mu}
\ee
in the amplitudes. 

Computing the density of eigenvalues, one is able to evaluate the
partition function and free energy in a genus expansion. Correlators
of observables can also be computed, as we will indicate in a
subsequent section.

This much of the discussion is valid for any compactification radius $
R$, including the limit $ R\to\infty$. But at $R=1$ we will see that
the formulae acquire special properties.

\subsection{Free Energy}

Because computations in matrix quantum mechanics are highly
technical, we will only quote some relevant results and refer the
reader to Ref.\cite{KlebanovReview} and references therein for the
derivations.

The free energy $\cF(\mu)\equiv\log\cZ(\mu)$ of the $ c=1$ string was
first obtained for arbitrary $R$ by Gross and Klebanov\cite{GroKle}:
\be
{\del^2\cF(\mu)\over\del\mu^2} = \hbox{Re} \int_0^\infty {dt\over t}
e^{- i\mu\, t}{ { t/2}\over \sinh  t/2}~
{  t/2R\over \sinh  t/ 2R}
\ee
This is to be understood as an asymptotic expansion in powers of 
$1/\mu^{ 2}= g_{ s}^{ 2}$. The dependence on the compactification
radius $R$ arises from the last factor in the integrand. In the limit
$R\to\infty$, this factor goes to 1.

The above formula can be integrated twice in $\mu$ to find the
 free energy.  The integration constants are  non-universal
terms  that will be unimportant.

Performing the expansion in $1/\mu^2$, we find:
\be
{\del^2\cF(\mu)\over\del\mu^2}= -\log\mu +\sum_{ g=1}^{ \infty}
{f_{ g}(R)\over (4R)^{ g}}\,\mu^{ -2g}
\ee
where
\be
f_{ g}(R) = (2g-1)! \sum_{ k=0}^{ g} |2^{ 2k}-2| |2^{
2(g-k)}-2| {|B_{2k}|\over (2k)!}  {|B_{2(g-k)}|\over (
2(g-k))!}\,R^{g-2k} 
\ee
and $B_{ 2k}$ are the  Bernoulli numbers.

Something special happens at $R=1$. One can use a 
bilinear identity on Bernoulli numbers, due to Gosper\footnote{For just about
everything you need to know about Bernoulli numbers, including this
identity, see Ref.\cite{Bernoulli}.}:
\be
\sum_{i=0}^n {(1-2^{1-i})(1-2^{i-n+1})B_{n-i}B_i\over
(n-i)!i!} = {(1-n)B_n\over n!}
\ee
Using this identity, we get:
\be
{\del^2\cF(\mu)\over\del\mu^2}\Big|_{ R=1} =
-\log\mu + \sum_{ g=1}^{\infty} {2g-1\over 2g} |B_{
2g}|\,\mu^{ -2g}
\ee

Integrating, the free energy is found to be:
\be
\cF(\mu)_{ R=\!1} ~=~ \sum_{ g=0}^{\infty} {|B_{ 2g}|\over
2g(2g-2)}\,\mu^{ 2-2g}
\ee
For genus $g=0,1$ the coefficients are formally
divergent (because of the dropped integration constants).

It is remarkable that the above expression arises from doing matrix
quantum mechanics in the double scaling limit. As we will see, it is
closely related to the (virtual) Euler characteristic of the moduli
space of Riemann surfaces. This is the first of many special
properties of $c=1,R=1$ that we will encounter.

\subsection{Tachyon correlators}

Let us first consider the zero-momentum tachyon.
The corresponding operator $T_0= e^{2\phi}$ is called the 
cosmological operator, because it is  conjugate to the
cosmological constant $\mu$:
\be
\langle T_0\, T_0\,\cdots\, T_0\rangle_{ g} ~\sim~ {1\over s!}{\del^{
s}\over
\del\mu^s}\cF(\mu)
\ee

Next consider $T_k$ for general nonzero $k$.  At $R=\infty$, $k$ is a
continuous real variable, while at $ R=1$, it is quantised to be a
(positive or negative) integer.  Because $X$ is a free field, the
total $k$ is conserved: 
\be 
\langle T_{ k_1}\, T_{ k_2}\,\cdots\, T_{ k_n}\rangle_{ g} ~=~0\quad
\hbox{unless}\quad \sum_{ i=1}^{ n} k_{ i}=0
\ee

In matrix quantum mechanics, one must identify the operators that
correspond to the massless tachyon modes. This procedure is less
straightforward than computing the BRST cohomology in the continuum
case. It turns out that the correct local operators are defined as
moments of loop operators\cite{Moore,MooSei}, up to normalisation --
about which we will say more below. Define:
\be
\cO(l,k) = \int dx\, e^{ikx}\tr\, e^{-lM(x)} ~\sim~ l^{
k}\, P_{ k} + \cdots
\ee
where we keep the leading term for small $ l$.

Correlators of the $P_{ k}$ the follow by computing correlation
functions of the eigenvalue density. For general $R$, these
computations were performed in Refs.\cite{KleLow,DijMooPle}. For
example, the three point function is given by:
\bea
&&\langle P_{ k_1}\,P_{ k_2}\, P_{ k_3}\rangle  
= R\,\delta(k_1+k_2+k_3)\,\mu^{\half(|k_1|+|k_2|+|k_3|-2)}
\Gamma(1-|k_1|)\Gamma(1-|k_2|)\Gamma(1-|k_3|)\times\nonumber\\[3mm]
&&\Bigg[1- {1\over 24R}(|k_3|-1)(|k_3|-2)
\left\{R(k_1^2 + k_2^2
-|k_3|-1)-{1\over R}\right\}\,\mu^{-2} + \cO(\mu^{-4})\Bigg]
\eea
It is a remarkable achievement of matrix quantum mechanics  that 
this result is actually known to all orders in $1/\mu^2$. Despite
appearances, it is totally symmetric under interchange of
$k_1,k_2,k_3$, and also invariant under $k_i\to -k_i$ as required by
translational symmetry in the $X$ coordinate.

By contrast, amplitude computations in the continuum Liouville theory
have only been done for genus $g=0$\cite{DifKut}. The three-point
function in this formulation was found to be:
\be
\langle T_{ k_1}\,T_{ k_2}\, T_{ k_3}\rangle_{ g=0} 
= R\,\delta(k_1+k_2+k_3)\,\mu^{\half(|k_1|+|k_2|+|k_3|-2)}
{\Gamma(1-|k_1|)\over \Gamma(|k_1|)}
{\Gamma(1-|k_2|)\over \Gamma(|k_2|)}
{\Gamma(1-|k_3|)\over \Gamma(|k_3|)}
\ee
It follows that the  continuum tachyons $T_{ k}$ and the  matrix
model puncture operators $P_{ k}$ are related by:
\be
T_{ k} = {1\over \Gamma(|k|)}P_{ k}
\ee
At  zero momentum  this represents an  infinite  change of
scale between the two descriptions\footnote{This is responsible for
the fact that the cosmological operator is not really $e^{2\phi}$
but $\phi e^{2\phi}$. Indeed, we see that
\be
P_0 = \lim_{k\to 0}{T_k\over k}\sim -\phi e^{2\phi} + \hbox{infinite
term}
\ee}. For other values of $k$
it is just a  finite  change.

We see that whenever $k$ is a nonzero integer, the amplitudes for both
$T_{ k}$ and $P_{ k}$ develop poles. Integer momenta are precisely the
ones for which additional discrete states (to which we briefly
referred in the Introduction) exist in the theory. The divergences in
tachyon amplitudes are believed to be due to the production of these
states in intermediate channels.

At $ R=1$, since $ k$ is always an integer,  every
correlator  of $T_{ k}$ or $P_{ k}$ (except for $P_0$)
is  divergent. Hence it is necessary to define  ``amputated tachyon
operators'' $\cT_{ k}$:
\be
\cT_{ k} ~=~ {\Gamma(1+|k|)\over \Gamma(1-|k|)}\,T_{ k} ~=~ 
{k\over \Gamma(1-|k|)}P_{ k}
\label{amputated}
\ee
The correlators of these operators have no poles\footnote{In the
Minkowski theory these leg pole factors are phases. Their
physical role has been discussed by Polchinski\cite{PolConsist}.}. 
Indeed, the three-point function of amputated
tachyons is:
\bea
\langle \cT_{ k_1}\,\cT_{ k_2}\, \cT_{ k_3}\rangle &&
= \delta(k_1+k_2+k_3)\,
\mu^{\half(|k_1|+|k_2|+|k_3|-2)}\,
|k_1\,k_2\,k_3|\times\nonumber\\[3mm]
&&\hskip -1cm\Bigg[1- {1\over 24}(|k_3|-1)(|k_3|-2)
(k_1^2 + k_2^2
-|k_3|-2)\,\mu^{-2} + \cO(\mu^{-4})\Bigg]
\eea

The absence of poles in the amplitudes for amputated tachyons suggests
that we are really working with a somewhat different theory, in which
the other discrete states are no longer present. The discrete tachyons
appear to be the only surviving states. In particular, this means that
the radius perturbation, which should be turned on to continuously
deform $R$ away from its self-dual value, is absent. In this sense,
the theory is ``stuck'' at $R=1$\footnote{One can take orbifolds to
get other integer values of $R$.} The above properties characterise the
topological $c=1$ background\footnote{Note that ``winding tachyons'',
which are winding modes around the $X$ direction, should still exist,
though they have not yet been satisfactorily understood in topological
matrix models.}. 

It is intriguing that the coefficient of the $\mu^{-2}$ term is an
integer for every $k_1,k_2,k_3$ satisfying $k_1+k_2+k_3=0$. Explicit
computation of higher-order terms in this and other correlators
suggests that this is always the case, though there does not seem to
be an explicit statement or proof in the literature. If every term is
an integer, this would certainly be special to $R=1$, as this property
can be seen not to hold for any other $R$.

An exact generating function for all discrete-tachyon
correlators, in every genus, is known\cite{DijMooPle}. This will be
described in a subsequent section.

\section{Riemann Surfaces and the Penner Matrix Model}\label{riemannpenner}

As has been indicated, the above results were derived starting from
matrix quantum mechanics, using the free fermion description. The first
indication that these are related to the topology of Riemann surfaces
arose in the study of the Penner matrix model\cite{Penner}, via the
work of Distler and Vafa\cite{DisVaf}.
To explain this model, we must first discuss some mathematical issues
relating to the moduli space of Riemann surfaces of genus $g$.  

\subsection{Moduli space of Riemann surfaces and its topology}

The moduli space of a compact Riemann surface of genus $g$ without
punctures, $\cM_g$, is a manifold of complex dimension $3g-3$ (for
$g>1$).  It arises as the quotient of the Teichm\"uller space by the
action of the mapping class group. Since this action has fixed points,
the moduli space $\cM_g$ has orbifold-like singularities, 

What is the simplest topological invariant of $\cM_g$?  For a
$D$-dimensional smooth manifold, we can define the the {\em Euler
characteristic} $\chi$ by making a simplicial decomposition ${\cal S}$
of the manifold and then evaluating:
\be
\chi = \sum_{i\in {\cal S}} (-1)^{d_i}
\ee
where $d_i$ is the dimension of the $i$th simplex, and the sum is over
all the simplices in the complex ${\cal S}$.

In the presence of orbifold singularities, the natural quantity to
define is the {\em virtual Euler characteristic} $\chi_V$. Here each
term in the sum is divided by the order of a discrete group $\Gamma_i$
that fixes the $i$th simplex. Thus:
\be
\chi_V = \sum_{i\in{\cal S}} {(-1)^{d_i} \over \#(\Gamma_i)}
\label{vecdef}
\ee
It was found by Harer and Zagier\cite{HarZag} that for the moduli
space of unpunctured Riemann surfaces, the virtual Euler
characteristic is:
\be
\chi_{ g} =  {B_{ 2g}\over 2g(2g-2)}
\ee
where $ B_{ 2g}$ are the Bernoulli numbers.

One can also consider the moduli space $\cM_{g,s}$ of Riemann surfaces
of genus $g$ with $s$ punctures. This has complex dimension $3g-3+s$
(for $g>1$, or $g=1,s\ge 1$, or $g=0,s\ge 3$). Its virtual Euler
characteristic is given by:
\be
\chi_{ g,s} = {(-1)^{s} (2g-3+s)!(2g-1)\over (2g)!\, s!} B_{
2g} = (-1)^s \pmatrix{2g-3+s\cr s}\chi_{ g}
\ee

The above results for $\chi_{g}$ and  $\chi_{g,s}$
are related in  a suggestive way. In string
theory, one expects that $s$ punctures can be created by differentiating the
vacuum amplitude $s$ times with respect to the cosmological constant. 
And indeed, as noticed by Distler and Vafa\cite{DisVaf}, one has the identity:
\be
{1\over s!} {\del^s\over \del\mu^s} \left(\chi_g  \mu^{2-2g}\right)
= \chi_{g,s}\,\mu^{2-2g-s}
\ee
Thus, if we could invent a model whose genus $g$ partition function is
$\chi_g$, we might expect it to bear some relation to string
theory. Below we will see that this is indeed the case.

\subsection{Quadratic differentials and fatgraphs}

The above results were obtained by triangulating the moduli space of
punctured Riemann surfaces. Such a triangulation was constructed
by Harer\cite{Harer} in terms of {\em quadratic differentials}, 
using a theorem due to Strebel\cite{Strebel}. It is instructive to
sketch how this was done.

On any Riemann surface with a finite number of marked points, one can
define a meromorphic quadratic differential
\be
\eta = \eta_{zz}(z)\,dz^2
\ee
which has poles at the locations of the marked points. This can
equivalently be considered as a holomorphic differential on the
punctured Riemann surface, which has these marked points removed.

Under a change  of coordinates:
\be
z\to z'(z)
\ee
a quadratic differential transforms as:
\be
\eta'_{z'z'}(z') = \left({\del z\over\del z'}\right)^2\eta_{zz}(z)
\ee
This differential can be used to invariantly define the length of a
curve $\gamma$ on the Riemann surface. The length  is simply defined to
be:
\be
|\gamma|_\eta = \int_\gamma \sqrt{|\eta(z)|}|dz|
\label{metric}
\ee
Indeed, defining a new coordinate via
\be
dw= \sqrt{\eta(z)}dz
\ee
we see that this length is the ordinary length of the curve in the
Euclidean sense, in the $w$ coordinate.

Now a quadratic differential is not in general unique, but it was
shown by Strebel\cite{Strebel} that with some extra properties, such a
differential is unique upto multiplication by a positive real
number. For this, let us consider a geodesic curve under the metric
defined above. At any point, such a curve will be called {\em
horizontal} if $\eta$ is real and positive along it, and {\em
vertical} if $\eta$ is real and negative. The horizontal curves define
flows along the Riemann surface.

The flow pattern is regular except at zeroes and poles of $\eta$. Here
the flows exhibit interesting properties. At an $n$th-order zero of
the quadratic differential, precisely $n+2$ horizontal curves meet at
a point. To see this, let us write down what the differential looks
like near this zero and along the radial direction:
\be
\eta\sim z^n(dz)^2 \sim e^{i(n+2)\theta}dr^2
\ee
From this we see that as we encircle the zero, there are precisely
$n+2$ values of the angle $\theta$ at which this differential is
positive, or horizontal.

At a double pole, if the coefficient is real and negative, the flows
form concentric circles around the point. This follows from the fact
that near the pole, and along the angular direction, the differential
looks like:
\be
\eta\sim -c{dz^2\over z^2}\sim c\, d\theta^2
\ee
Thus, in the $\theta$ direction, the differential is positive, 
or horizontal, at all points surrounding the double pole.

Other behaviours are possible at poles of order
$n=1$ or $n\ge 3$, or if the coefficient of $\eta$ at a double pole is
complex (for a comprehensive description, see Figs.5-13 of
Ref.\cite{Strebel}). But we do not need to consider these other
behaviours because we will restrict our quadratic differentials not to
have such poles.

In fact, we will specialise to a class of quadratic differentials
which have a double pole at a point $P$, at which the coefficient $c$
is required to be real and negative. Next, we require that all
smooth horizontal trajectories (i.e., those that do not pass through
zeroes of $\eta$) form closed curves. Quadratic differentials
satisfying all these conditions exist, and are called {\em
horocyclic}. 

An illustration of the flow pattern associated to a horocyclic
quadratic differential is given in Fig.2. Note that the vertex has
five lines meeting at a point, indicating a third-order zero (the
differential also has additional zeroes not shown in the diagram).

\EPSFIGURE{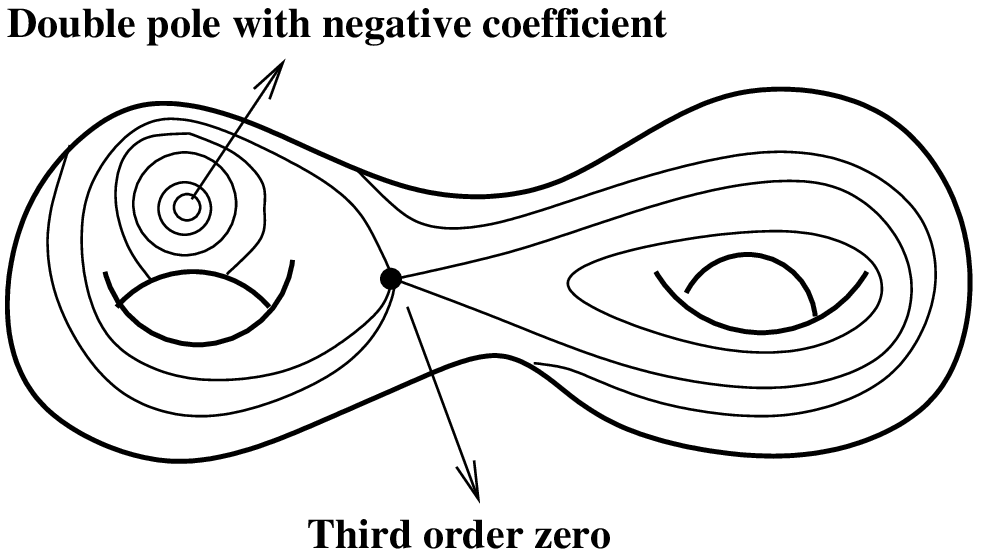}{Riemann surface with the 
flow pattern of a horocyclic quadratic differential.}
\vspace*{2mm}

Strebel's theorem states that on every Riemann surface of genus $g$
with 1 puncture, for fixed values of its moduli, there exists a
{\em unique} horocyclic quadratic differential with a double pole at
the puncture. (The uniqueness is upto multiplication by a real positive
number).

Thus, by studying how these quadratic differentials vary as we vary
the moduli, we get information about the moduli space $\cM_{g,1}$ of a
once-punctured Riemann surface. In view of some mathematical results
relating the moduli spaces of Riemann surfaces of a fixed genus $g$
but different numbers of punctures $s$\cite{Harer}, this construction
actually enables us to study the moduli space $\cM_{g,s}$ of
$s$-punctured Riemann surfaces as well.

The main motivation for having gone through this mathematical
description here is that we will now see the emergence of
``fatgraphs'', otherwise known as 't Hooft diagrams or large-$N$
diagrams.  Most of the flows are closed and smooth, but there are
singular ones that branch into $n+2$-point vertices at $n$th order
zeroes of $\eta$. We can think of these singular flows as defining a
Feynman diagram, whose vertices are the branch points. Each double
pole of $\eta$ is a point around which the flows form a loop, hence
the number of loops of the diagram is the number of double poles,
which is the number of punctures of the original Riemann
surface. Finally, because the flows that do not pass through a zero
are smooth, each singular flow can be ``thickened'' into a smooth
ribbon in a unique way, and we arrive at a fatgraph.

The fatgraphs with a single face triangulate the moduli space
$\cM_{g,1}$ in the following way. Consider the lengths of each edge of
a fatgraph, as computed in the metric defined in Eq.(\ref{metric})
above. We have an overall freedom of scaling the whole Riemann
surface, which clearly does not change the moduli. So to vary the
moduli, we must change the lengths of the different edges keeping the
total length fixed. By Strebel's correspondence between quadratic
differentials and moduli, this sweeps out a region of the moduli space
of the Riemann surface. The dimensionality of this region will be
$E-1$ where $E$ is the number of edges of the graph.

As Harer argues, this region swept out will not be the whole moduli
space, but a simplex of it. In a simplicial decomposition, at the
boundary of a simplex we find a lower-dimensional simplex. In terms of
fat graphs, a boundary occurs whenever a length goes to zero and two
vertices meet. As an example, when two three-point vertices meet, we
obtain a four-point vertex. The new graph thus obtained sweeps out a
lower-dimensional simplex in the moduli space as its remaining lengths
are varied, and this process continues.

Now the virtual Euler characteristic can be defined directly in terms
of fatgraphs. We consider the set of all fatgraphs of a given genus
$g$ and a single puncture. Let us call the set ${\cal S}$ and label
each distinct graph by an integer $i\in{\cal S}$. Let $\Gamma_i$ be
the automorphism group of a fatgraph. Then, defining $d_i = E-1$, we
write
\be
\chi_V = \sum_{i\in{\cal S}} {(-1)^{d_i} \over \#(\Gamma_i)}
\label{vecfat}
\ee
which is analogous to Eq.(\ref{vecdef}) above, except that now
the sum is over fatgraphs rather than over simplices. In particular,
the automorphism group of the fatgraph is the same as the group that
fixes the corresponding simplex. Very low-dimensional simplices do not
contribute to $\chi_V$ since at some point the corresponding fatgraph
has an automorphism group of infinite order, in which case the
denominator of Eq.(\ref{vecfat}) becomes infinite and the
corresponding term vanishes.

Let us check how the correspondence between fatgraphs and quadratic
differentials works out in practice. The fatgraphs we have been
considering have $V$ vertices, $E$ edges and 1 face. These integers satisfy:
\be
V-E + 1 = 2-2g
\ee
where $g$ is the genus of the Riemann surface on which the graph is
drawn.

We also have the relations
\be
V=\sum_k v_k,\quad E=\half \sum_k kv_k
\label{edgerel}
\ee
where $v_k$ is the number of $k$-point vertices. From these relations,
we get:
\be
\sum_k (k-2)v_k = 4g-2
\ee
All integer solutions of this equation, i.e. all choices of the set
$\{v_k\}$ for fixed $g$, are valid graphs that correspond to
simplices in the triangulation of $\cM_{g,1}$.

Let us recast the above equation as
\be
\sum_k (k-2)v_k -2 = 4g-4
\ee
Then, noticing that $k-2$ is the order of the zero represented by a
$k$-point vertex, we see that the first term on the left is the total
number of zeroes (weighted with their multiplicity) of the quadratic
differential corresponding to the given fatgraph. Moreover, the
differential has precisely one double pole, so the second term is
minus the (weighted) number of poles.  Thus this result is in accord
with a theorem which states that for meromorphic quadratic
differentials on a Riemann surface of genus $g$,
\be
\#(\hbox{zeroes}) -\#(\hbox{poles}) = 4g-4
\ee

A particular solution that is always available is $v_3=V$, $v_k=0,k\ge
4$. Clearly this gives the maximum possible number of vertices and
therefore also edges. In this case,
\be
V = 4g- 2
\ee
and we see from Eq.(\ref{edgerel}) that the number of edges is
\be
E= {3\over 2}V = 6g-3 
\ee
Thus the dimension of the space spanned by varying the lengths of the
graph keeping $s$ lengths fixed, is:
\be
E-1 = 6g-4
\ee 
which is the real dimension of $\cM_{g,1}$. Thus the graphs with only
cubic vertices span a top-dimensional simplex in moduli space. All
other graphs arise by collapse of one or more lines, merging two or
more 3-point vertices to create higher $n$-point vertices. These
correspond to simplices of lower dimension in the moduli space.

As an example, in genus 3 with one puncture, the moduli space has real
dimension 14. Hence the graph that describes the ``bulk'' of moduli
space should have 15 edges and 10 3-point vertices. Also, because
there is a single puncture, the graph should have just one face. 
In summary, we have
\be
V=10,~ E= 15,~ s=1\quad\Rightarrow\quad V-E+s=-4 = 2-2g
\ee
as expected.

\subsection{The Penner model}\label{pen}

In 1986, Penner\cite{Penner} constructed a matrix model that provides
a generating functional for $\chi_{ g,s}$. The Penner model is defined
in terms of $N\times N$ random matrices whose ``fatgraphs'' are
precisely the ones described in the previous subsection. 
The free energy $\cF=\log \cZ$ of this model then has
the expansion:
\be
\cF = \sum_{ g} \cF_{ g} = \sum_{ g,s}~\chi_{ g,s}~N^{
2-2g}~ t^{ 2-2g-s}
\label{freeenergy}
\ee
where $t$ is a  parameter  of the model. The term $s=0$
is not present in the sum.

Let us first present the model and then show why it is correct. The
model is given by an integral over Hermitian random matrices:
\bea
\cZ &=& \cN_P\int\,[dQ]\, e^{\ts -Nt\tr \sum_{k=2}^\infty { 1\over
 k}\,Q^{ k}}\label{penmod}\\
&=& \cN_P\int\,[dQ]\, e^{\ts Nt\tr \big(\log(1-Q) + Q\big)}
\eea
where $\cN_P$ is a normalisation factor given by:
\be
\cN_P^{ -1} = \int\,[dQ]\, e^{\ts -Nt\tr { 1\over  2}Q^{ 2}}
\label{pennernorm}
\ee
and the matrix measure $[dQ]$ is given by:
\be
[dQ] \equiv \prod_i dQ_{ii}\prod_{i<j}dQ_{ij}\,dQ^*_{ij}
\ee
This action has  all powers $k\ge 2$ of the random matrix
appearing in it. The model is to be considered as a perturbation
series around $Q\sim 0$.

To show that this model is correct, we must show that its fatgraphs
are in one to one correspondence with those arising from quadratic
differentials as discussed in the previous section. Thus we must show
that the partition function involves a sum over all (connected and
disconnected) fatgraphs for a fixed genus $g$ and number of faces $s$,
multiplied by the weighting factor
\be
{(-1)^{E-s}\over\#(\Gamma_i)}\,N^{2-2g} t^{2-2g-s} = {1\over
\#(\Gamma_i)}\,
(-Nt)^V (Nt)^{-E}(N)^s 
\ee
By taking the logarithm of the partition function we end up
with the sum over connected graphs in the familiar way.

In the above expression $\Gamma_i$, the automorphism
group, is the collection of maps of a given fatgraph to itself such that
(i) the set of vertices is mapped onto itself, (ii) the set of edges
is mapped to itself, and (iii) the cyclic ordering of each vertex is
preserved. A key result due to Penner\cite{Penner} is that the order of this
group is equal to:
\be
\#(\Gamma_i) = {1\over V!}\prod_{\hbox{vertices}}{1\over k} \times C
\ee
where $C$ is the combinatoric factor labelling how many distinct
contractions lead to the same graph. Now this is exactly the factor
that arises if we obtain our diagrams as the expansion of the matrix
integral Eq.(\ref{penmod}), with the $V!$ coming from the order of
expansion of the exponent, the ${1\over k}$ factors being built into
the action, and the combinatoric factor being the usual one.

The remaining factors are straightforward. Each vertex has a factor
$-Nt$, each edge is a propagator given by $(Nt)^{-1}$, and each face
gives the usual index sum $N$. This completes our informal derivation
of the Penner model\footnote{Another derivation, as well as
proposed generalisations, can be found in Ref.\cite{Chaudhuri}.}.

\subsection{Double-scaled Penner model}

In principle, the Penner model has nothing to say about the moduli
spaces $\cM_{g,0}$ of {\em unpunctured} Riemann surfaces. As we have
noted, the expansion of the partition function does not contain terms
with $s=0$.

However, it was noticed by Distler and Vafa\cite{DisVaf} that in a
suitable double-scaling limit, the model seems to describe unpunctured
Riemann surfaces. Their observation goes as follows. Start with the
equations:
\bea
\cF &=& \sum_{ g} \cF_{ g} = \sum_{ g,s}~\chi_{ g,s}~N^{
2-2g}~ t^{ 2-2g-s}\\
\chi_{ g,s} &=& {(-1)^{ s} (2g-3+s)!(2g-1)\over (2g)!\, s!} B_{
2g} = (-1)^s \pmatrix{2g-3+s\cr s}\chi_{ g}
\eea
where we have inserted the known value of $\chi_{g,s}$.
 
Let us work with genus $g>1$. It is straightforward to explicitly perform the sum over
$s$, and one gets:
\be
\cF_{ g} = {B_{ 2g}\over 2g(2g-2)} (Nt)^{ 2-2g}\Big(
(1+t)^{ 2-2g} -
1\Big) 
\ee
Now Distler and Vafa took the limit $N\to\infty$ and $t\to t_{ c}=-1$, 
keeping fixed the product $N(1+t)=-\nu$. This leads immediately to the result:
\be
\cF_{ g} = {B_{ 2g}\over 2g(2g-2)} \,\nu^{ 2-2g} = \sum_g \chi_g \nu^{2-2g}
\ee
and we have recovered the virtual Euler characteristic of $\cM_{g,0}$!

Thus the Penner model, originally designed to study the moduli space
of punctured Riemann surfaces, describes unpunctured ones too -- but
only in the special double-scaling limit above.
We can remove some of the mystery of the double-scaling limit by
defining, for any finite $N$ and $t$, the parameter 
\be
\nu = -N - Nt
\ee
in terms of which the matrix model can be written:
\be
\cZ = \cN_P\int\,[dQ]\, e^{-(\nu+N)\ts\tr \big(\log(1-Q) + Q\big)}
\ee
Now the double-scaling limit is simply the ordinary large-$N$ limit,
$N\to\infty$, with $\nu$ held fixed.

From the discussions of the previous sections, it should be evident that
the double-scaled Penner model is closely related to $c=1,R=1$. 
Its free energy:
\be
\cF_P(\nu) = \sum_{ g} {B_{ 2g}\over 2g(2g-2)} \,\nu^{ 2-2g}
\ee
 is  {\em almost} identical  to the free energy of
the $c=1,R=1$ string theory:
\be
\cF_{c=1}(\mu) = \sum_{ g} {|B_{ 2g}|\over 2g(2g-2)} \,\mu^{ 2-2g}
\ee
However, there is an important issue of signs. It is well-known that the
Bernoulli numbers alternate in sign:
\be
|B_{ 2g}| =  (-1)^{ g-1} B_{ 2g}
\ee
Therefore if we define $\nu=-i\mu$, we can write:
\be
\cF_P(-i\mu) = \sum_{g=0}^\infty {|B_{ 2g}|\over
2g(2g-2)}\,\mu^{ 2-2g} = \cF_{c=1}(\mu)
\ee
Thus the double-scaled Penner model (at {\em imaginary} cosmological
constant) has the same free energy as the $ c=1,R=1$ string. This
intriguing correspondence calls for an explanation.

In the last section we will try to argue that the double-scaled Penner
model could be related to the Euclidean $c=1,R=1$ theory, and could
perhaps be derived from it, as the matrix integral for $N$ D-instantons of the
latter theory. If true, this might both explain the above coincidence
and also provide a role for D-instantons of the noncritical $c=1$
string. 

\section{The $W_\infty$ Matrix Model}\label{winf}

Let us now return to the solution of the $c=1,R=1$ matrix quantum
mechanics. We will now focus on its tachyon amplitudes.

\subsection{$c=1$ amplitudes and $W_\infty$}

As was already mentioned, for this theory a generating functional for
all tachyon correlators to all genus has been
obtained\cite{DijMooPle}. This functional $\cF(t,\tbar)$ depends on an
infinite set of couplings $t_k,\tbar_k$ such that:
\be
\langle \cT_{ k_1}\,\ldots\, \cT_{ k_n}\,
\cT_{ -l_1}\,\ldots\,\cT_{ -l_m}\rangle
= {\del\over \del t_{ k_1}}\cdots {\del\over
\del t_{ k_n}}
{\del\over \del \tbar_{ l_1}}\cdots
{\del\over \del \tbar_{ l_m}}\, \cF(t,\tbar)\Bigg|_{ t=\tbar=0}
\ee
where on the LHS we have  connected  amplitudes. We see that the
couplings $t_k$ are sources for tachyons of positive momentum $k$,
while the couplings $\tbar_k$ are sources for tachyons of negative
momentum $-k$. These tachyons are ``amputated'' in the sense of
Eq.(\ref{amputated}).

Instead of constructing the generating functional directly in terms of
$t_k,\tbar_k$, it turns out necessary to encode the parameters $t_k$
into a constant $N\times N$ matrix $A$.  The $t_k$ are defined in
terms of $A$ by the relation:
\be
t_{ k} = {1\over \nu k}\tr A^{ -k}
\label{kontsmiwa}
\ee
which is sometimes called the Kontsevich-Miwa transform.  As
$N\to\infty$, this matrix can encode infinitely many independent
parameters $t_{ k}$.

Now using matrix quantum mechanics at $R=1$, it was
shown\cite{DijMooPle} that $Z(t,\tbar)= e^{\cF(t,\tbar)}$ is a
$\tau$-function of the Toda hierarchy, satisfying an infinite set of
constraints that form a $W_\infty$ algebra. These constraints were
subsequently rewritten in the following form\cite{ImbMuk}, which is
the version that we will need for the subsequent discussion:
\be 
{1\over (-\nu)}{\del\cZ\over \del \tbar_{ n}} = {1\over(-\nu)^{
n}}(\det A)^{\nu} \tr
\left({\del\over\del A}\right)^{ n} (\det A)^{-\nu}\,\cZ(t,\tbar)
\ee
Here $\nu=-i\mu$, and $\mu$ is the cosmological constant. The
correlators determined by the above expression are actually invariant
under $k_i\to-k_i$, though this is far from manifest, since $t_n$ and
$\tbar_n$ do not appear symmetrically.

The $W_\infty$ constraints can be integrated to give a matrix
model\cite{ImbMuk} as follows\footnote{The same constraints were
integrated in Ref.\cite{DijMooPle} but owing to some technical errors,
as explained in Ref.\cite{ImbMuk}, the resulting matrix model obtained
in Ref.\cite{DijMooPle} was not correct.}.  Let us start by assuming
that $\cZ(t,\tbar)$ is an integral over constant matrices $M$:
\be
\cZ(t,\tbar) = (\det A)^{\nu} \int [dM]\,e^{\ts\tr V(M,A,\tbar)}
\label{constint}
\ee
where $[dM]=\prod_i dM_{ ii}\prod_{i<j} dM_{ ij}dM^*_{ ij}$.
The potential $ V$ is determined by  imposing the above differential
equation:
\be
\left[{1\over (-\nu)}{\del\over \del\tbar_{ n}} - {1\over
(-\nu)^{ n}}\tr\left({\del\over \del A}\right)^{ n}\right] \,\int
[dM]\,e^{\ts\tr V(M,A,\tbar)} =0
\ee
This determines:
\be
V(M,t,\tbar) = -\nu\Big(MA + \sum_{ k=1}^{\infty} \tbar_{ k}
M^{ k}\Big) + f(M)
\label{potential}
\ee
where $f(M)$ is a function independent of $A,\tbar$. 

The function $f(M)$ can be determined using a boundary condition, arising
from conservation of the tachyon momentum. Momentum conservation 
tells us that if we set all the $\tbar_k$ equal to zero, then $Z(A,0)$
has to be independent of $A$.  From Eqns.(\ref{constint}) and
(\ref{potential}), we get:
\be
Z(t,0) = (\det A)^{\nu} \int [dM]\, e^{\ts-\nu \tr MA + \tr f(M)}
\ee
Upon changing variables $ M\to M A^{-1}$, we see that:
\be
[dM]\to (\det A)^{-N}[dM]
\ee
It follows that:
\bea
Z(t,0) &=& (\det A)^{\nu-N} \int [dM]\, e^{\ts-\nu \tr M + \tr
f(MA^{-1})}\\
&=& \int [dM]\, e^{\ts-\nu \tr M + \tr
f(MA^{-1}) + (\nu-N)\tr\log A}
\eea
This  uniquely  determines the function $f(M)$ to be:
\be
f(M) = (\nu-N)\log M
\ee

Putting everything together, we see that the generating function 
of all tachyon
amplitudes in the $c=1,R=1$ string theory is:
\bea
\cZ(t,\tbar) &=& (\det A)^{\nu} \int [dM]\, e^{\ts\tr(-\nu MA
+(\nu-N)\log M
-\nu\sum_{k=1}^{\infty} \tbar_{ k} M^{ k}) }\label{firstform}\\
&=& \int [dM]\, e^{\ts\tr(-\nu M
+(\nu-N)\log M
-\nu\sum_{k=1}^\infty \tbar_{ k} (MA^{-1})^{ k} )}
\label{winfty}
\eea 
This is the $W_\infty$ matrix model\cite{ImbMuk}.
The second form was obtained by redefining $M\to MA^{-1}$.
We also recall that 
\be 
t_{ k} = {1\over \nu\, k}\tr A^{-k} 
\ee

Note that for the matrix integral to be well-defined, 
the matrix $M$ (more precisely, its  eigenvalues) 
must be  positive semi-definite.

The $W_\infty$ model has a number of interesting properties. One of
these is the {\em puncture equation}, which says that the shift
\be
A\to A-\epsilon I,\quad \tbar_1 \to \tbar_1 + \epsilon
\ee
changes the free energy $\cF$ by a simple additive factor:
\be
\cF(A-\epsilon I,\tbar_k + \delta_{k,1}\,\epsilon) = \cF(A,\tbar)
-\nu^2\sum_{k=1}^\infty \epsilon^k\,t_k 
\ee
This is easiest to see when the model is written in the form
Eq.(\ref{firstform}). We see that there are {\em two} linear terms in
$M$, one multiplying $A$ and the other multiplying $\tbar_1$. Their
variations under the above transformation cancel out, leaving only the
effect of varying the determinant factor in front of the
integral. Taking logarithms gives the above behaviour of the free
energy. 

Another property of the $W_\infty$ model is that it is invariant under
the simultaneous rescaling $t_k\to \lambda^{-k} t_k$ and $\tbar_k \to
\lambda^k\tbar_k$, for any (in principle, complex) parameter
$\lambda$. This is just momentum conservation. To see this invariance,
note first that by Eq.(\ref{kontsmiwa}, the rescaling $t_k\to
\lambda^{-k} t_k$ is the same as $A\to \lambda A$. Now the expression
Eq.(\ref{winfty}) is manifestly invariant under $A\to\lambda A,
\tbar_k \to \lambda^k\tbar_k$.

There have been other matrix models in the literature which have a
form similar to that of the $W_\infty$ model\footnote{An early attempt
to incorporate correlators in the Penner model is
Ref.\cite{ChaPan}.}. Some examples are the Chekhov-Makeenko
model\cite{CheMak}, the NBI matrix model\cite{FayMakOleSmiZar,AmbChe}
and the normal matrix model\cite{AleKazKos}. The first of these is
similar in structure to our $W_\infty$ matrix model but the
coefficients are different. The second one is proposed as a
modification of the IKKT matrix model by coupling it to a new matrix
with a log plus linear interaction.  The third example, the normal
matrix model, will be discussed in Sec. \ref{tachp}.

\subsection{Relation to the Penner model}\label{relpen}

The matrix integral of the $W_\infty$ model is convergent for real
positive $\nu$. Indeed, it is the matrix analogue of the
$\Gamma$-function. But we are presently interested only in its
perturbative expansion in inverse powers of $\nu^2$. If we send
$\nu\to -\nu$, we get the same perturbation series, even though the
integral is formally no longer convergent and has to be defined by
analytic continuation (like the $\Gamma$-function itself).

It turns out that this transformation is necessary in order to recover
the Penner model from the $W_\infty$ model. Carrying out the change
$\nu\to-\nu$, the resulting matrix integral is:
\be
\cZ(t,0) = \int [dM]\, e^{\ts\tr(\nu M
-(\nu+N)\log M)}
\label{kpunpert}
\ee
Now make the change of variables:
\be
M = \alpha\,(1-Q)
\label{kptopenner}
\ee
where $\alpha$ is a parameter to be determined.

Thus the $W_\infty$ matrix model is transformed into the following
matrix integral:
\be
\cZ = \cN \int [dQ]\, e^{\ts-(\nu+N)\tr\log(1-Q) -\nu\alpha\, Q}
\ee
where $\cN$ is a normalisation constant. 

Choosing $\alpha = 1+N/\nu$, we get the standard form of the
Penner model:
\be
\cZ= \cN \int [dQ]\, e^{\ts Nt\,\tr\big(\log(1-Q) +Q\big)}
\label{stdpenner}
\ee
where $t= -1 - \nu/N$. 

Notice that in the ``double-scaled'' limit of the Penner model, we
have $N\to\infty$ with $\nu$ finite, so the parameter $\alpha$ goes to
infinity. As a result, the Penner model perturbation series, around
$Q\sim 0$, actually corresponds to the $W_\infty$ model in the 
region of {\em large} $M$.

A word about normalisation is called for. The $W_\infty$ model determines
correlators of tachyons of nonzero integer momentum
(positive or negative). If insertions of $\cT_0$ are also required,
they can be obtained by differentiating in $\nu$. However, in the
derivation leading up to Eq.(\ref{winfty}), the overall
$\nu$-dependent normalisation was not fixed, as this does not
affect correlators for which at least one tachyon has nonzero momentum
(by momentum conservation, this means that at least two tachyons have
nonzero momentum).

This is why Eq.(\ref{stdpenner}) has a normalisation constant in front
of it. We could of course have at the outset chosen the normalisation
in Eq.(\ref{winfty}) to reproduce the correct normalisation factor of
the Penner model given in Eq.(\ref{pennernorm}). But amusingly, this
is not really necessary, for the following reason. We can extract
correlators of zero-momentum tachyons from the $W_\infty$ model by
taking a formal limit $k\to 0$ in the corresponding combinatorial
formulae. That this gives the right answer is demonstrated in
Ref.\cite{GhoImbMuk}.

\subsection{Relation to the Kontsevich model}

We have not so far discussed the original Kontsevich matrix
model\cite{Kontsevich}, but it is appropriate to make a brief mention
of it here. Like Penner's model, this model too was formulated to
solve a combinatoric problem, this time concerning stable cohomology
classes on moduli space. Via Witten's results relating these
cohomology classes to pure topological gravity and $c<1$
strings\cite{Witten}, this matrix model actually describes $c<1$
(rather than $c=1$) noncritical strings. 

The Kontsevich model is described by random matrices with the
following matrix integral:
\be
\cZ_K(t) = \int [dX]\, e^{\ts-\tr\half \Lambda X^2 +i\tr X^3/6}
\label{konts}
\ee
The $t_i$ on which the partition function depends are an infinite set
of parameters defined  by:
\be
t_k = -(2k-1)!!\,\tr \Lambda^{-2k-1}
\ee
Differentiating the free energy $\cF_K=\log \cZ_K$ with respect to the
$t_i$ gives rise to an infinite set of correlation functions of
operators that are analogous to the ``amputated discrete tachyons'' of
previous sections. In turn, these correlators are the amplitudes of a
class of $c<1$ string theories based on $(2,q)$ minimal models, the
ones originally studied as double-scaled matrix
models\cite{BreKaz,DouShe,GroMig}.

It was noticed by Kontsevich that his model is related to the matrix
Airy function:
\be
A(Y) = \int [d\hX]\, e^{\ts\, i\tr\left(\hX^3/3 - \hX Y\right)}
\label{airy}
\ee
Indeed, the change of variables:
\be
\hX = 2^{-1/3}X + Y^{1/2},\quad Y = - 2^{-2/3}\Lambda^2
\label{airychange}
\ee
brings Eq.(\ref{airy}) into the form of Eq.(\ref{konts}), with,
however, a significant prefactor depending on $\Lambda$. Thus the
perturbation series of one can be expressed in terms of the
perturbation series of the other.

It was also noted in Ref.\cite{Kontsevich} that the matrix Airy
function Eq.(\ref{airy}) is to be viewed as an asymptotic expansion at
large $Y$, for which the saddle point at $\hX\sim Y^{\half}$ dominates
and therefore the matrix $\hX$ is also large. On the other hand, the
corresponding perturbation series for the model of
Eq.(\ref{konts}) is around $X\sim 0$. This is strikingly similar
to the relation between the $W_\infty$ model of Sec. \ref{winf}, and the
Penner model of Sec. \ref{pen}. As noted in Sec. \ref{relpen}, the
former model expanded about large values of the matrix $M$ gets mapped
to the latter model expanded about $Q\sim 0$.

We can actually obtain the matrix Airy function directly as a special
case of the $W_\infty$ model of Eq.(\ref{winfty}). Starting
with Eq.(\ref{winfty}) at finite $N$, set $\nu=N$ and $\tbar_k =
\delta_{k,3}$. The result is:
\be
\cZ(t) = (\det\,A)^N \,\int dM~ e^{\ts-N\,\tr M A -N\tr M^3}
\ee
After a further rescaling of both $M$ and $A$, this is just
proportional to the integral Eq.(\ref{airy}). Thus the matrix Airy
function arises from the $W_\infty$ model by ``condensing'' a specific
negative-momentum tachyon, $t_3$.

Generalisations of the Kontsevich model to describe noncritical
strings based on $(p,q)$ minimal models do exist\cite{AdlMoe}. They
are related to generalisations of the matrix Airy integral where the
cubic term is replaced by a higher power. Clearly these too can be
obtained from the $W_\infty$ model by condensing, or turning on,
$\tbar_k$ for $k>3$.

For string theory amplitudes, we are interested not just in the matrix
integral but also the $A$-dependent or $\Lambda$-dependent
normalisations. So these observations do not show that $c<1$ strings
are obtained from the $c=1$ background by ``condensing'' discrete
tachyons, but they do seem to suggest this. Specifically, turning on
$\tbar_p$ or condensing $\cT_{-p}$ is related to the family of
$(p-1,q)$ minimal models coupled to gravity, for fixed $p$ and all
coprime $q$. More work is needed to precisely establish this
relationship.

\section{Liouville Matrix Model and D-Instantons}
\label{lastsec}

In this section we will rewrite the $W_\infty$ model in a form that
involves a Liouville-like potential. This form will be suggestive of a
D-instanton interpretation.

\subsection{Unperturbed $W_\infty$ and the Liouville Matrix Model}

Let us start with the model as written in Eq.(\ref{kpunpert}), where
all the $\tbar_k$ have been set to zero.  Notice that the coefficients
in the action have a very specific dependence. The linear term is
multiplied by a constant $\nu$ that remains fixed in the large-$N$
limit, while the log term has a coefficient $\nu-N$. Also, as we have
seen, the matrix $M$ has to be positive semi-definite. Both these
features look slightly unnatural in a matrix model, and suggest a
redefinition of variables. 

Define a new $N\times N$ matrix $\Phi$ by:
\be
M=e^{\ts\Phi}
\ee
Under this transformation, the matrix measure transforms as:
\be
[dM]\,(\det M)^{-N} = [d\Phi]
\ee
and the integral in Eq.(\ref{kpunpert}) becomes:
\be
\cZ = \int [d\Phi]\, e^{\ts-S} = \int [d\Phi]\, e^{\ts-\nu \tr\left(e^{\Phi}
-\Phi\right)}
\ee
where $[d\Phi]$ is an (analytically continued) matrix measure
appropriate for unitary matrices:
\be
[d\Phi] \equiv \prod_i d\phi_{i}\prod_{i<j}
\left(\sinh\half(\phi_i - \phi_j)\right)^2
\ee
Since $M$ is Hermitian and positive semi-definite, it follows that
$\Phi$ is Hermitian with no further restrictions. Thus we no longer
have to deal with a positivity constraint.  More remarkably, we have
also got rid of the explicit $N$ in the exponent. The matrix model in
the $\Phi$ variable is simply a matrix integral with a potential that
has a Liouville type exponential term, as well as a linear term.  We
will refer to it as the {\em Liouville matrix model}.

\subsection{Interpretation of the Model}

How do we interpret this model? The potential looks very much like
that on the worldsheet of a fundamental string in the $c=1$
background.  If we think of $\Phi$ as the Liouville field, the two
terms are reminiscent of the Liouville potential (cosmological
operator) and the linear dilaton potential respectively.

However, there is a different and more plausible interpretation.  The
coefficient of the entire matrix action is $\nu$, the cosmological
constant. In noncritical string theories this corresponds to the
inverse string coupling: $\nu = 1/g_s$. Thus the action of the
Liouville matrix model has a factor $1/g_s$ in front of it. This is
highly suggestive of a D-brane action. Since the matrices are
constant, the D-brane in question should be a D-instanton.

In critical superstring theory, a single D-instanton has essentially
no action corresponding to it. But $N$ D-instantons do have an action,
consisting of all the commutator terms that arise in the nonabelian
Yang-Mills action: \be S = {1\over g_s}\tr [A_\mu,A_\nu]^2 + \cdots
\ee Fermionic terms and higher-order commutators are also present.
All these commutator terms arise because the scalars transverse to the
instanton, and their fermionic partners, are promoted to matrices when
there are $N$ instantons. Systems of D-instantons for large $N$ are
associated to the IKKT\cite{IKKT} matrix model of type IIB string
theory.

Suppose we had $N$ D-instantons in $c=1$ string theory. What action
should we propose for them\footnote{I am grateful to Ashoke Sen for
discussions on this point.}?  One could try to address this by
constructing a boundary state and computing open-string
amplitudes. This approach would work only at weak coupling, that is
far out near the region of large negative Liouville field $\phi$. In
this region, the D-instanton will feel two forces, one due to the
cosmological term, i.e. the Liouville potential, which is an
excitation of the closed-string tachyon, and the other due to the
linear dilaton. The former would drive the brane towards weak
coupling, while the latter drives it to strong coupling. It is at
least plausible that the two competing forces have an equilibrium
position somewhere in the region of $\phi\sim 0$. However, it would be
hard to gather reliable information in that region, since the theory is
strongly coupled there.

Hence we leave this question for future work. What is quite striking,
though, is that without making any assumptions at all, we have
obtained a Liouville matrix model which precisely consists of a
competing exponential and linear potential. This might be considered
as evidence for a D-instanton interpretation of the $W_\infty$
matrix model.

This interpretation also adds something to the discussion in the
previous section on the relationship of the $W_\infty$ to the Penner
model. Recall that the transformation from the $W_\infty$ to the Penner
matrix, Eq.(\ref{kptopenner}) involves a large parameter $\alpha = 1 +
N/\nu$, such that at large $N$, small values of the Penner matrix
$Q\sim 0$ are mapped to large values of $M$. Now we see that in the
Liouville variable this is the region of large $\Phi$, or the strong
coupling region. This could be related to the fact that in the
D0-brane description of $c=1$ strings\cite{KleMalSei} via the ZZ
boundary state\cite{ZamZam}, the branes are thought of as being
localised at large $\phi$.

Some ingredients are clearly missing from the story. We did not find a
reason why the cosmological constant in this theory should be
imaginary. Also, a D-instanton action should depend on two additional
variables: the transverse scalar $X$ describing the $c=1$ direction,
and the open-string tachyon $T_{\hbox{\small open}}$. We will see that
something like $X$ does appear when we consider tachyon perturbations,
but we do not seem to find a role for $T_{\hbox{\small open}}$, unless
it is somehow ``mixed up'' with $\Phi$.

\subsection{Tachyon Perturbations}\label{tachp}

We proposed that the matrix model on $N$ D-instantons is equivalent to
the $W_\infty$ model without tachyon perturbations (i.e. at
$t,\tbar =0$). What happens when we turn on tachyon perturbations? 

If the proposal is correct, the D-instanton action should match
the full $W_\infty$ action as a function of $t,\tbar$:
\be
\cZ(t,\tbar) = \int [dM]\, e^{\ts\tr(-\nu M
+(\nu-N)\log M
-\nu\sum_{k=1}^\infty \tbar_{ k} (MA^{-1})^{ k} )}
\ee
where 
\be
t_{ k} = {1\over \nu\, k}\tr A^{-k}
\ee

Let us focus on the matrix $A$. If it is Hermitian then (for real
$\nu$) the $t_{ k}$ are real. But we could equally well choose $A$ to
be unitary, in which case the $t_{ k}$ are complex, since we only need
to differentiate $\cF(t,\tbar)$ near $t,\tbar=0$.

Let us therefore choose $A$ to be  unitary, and parametrise it as:
\be
A = e^{\ts iX}
\ee
where $X$ is Hermitian.

Now, in the Liouville-like variable $\Phi$, the $W_\infty$ matrix
model with tachyon perturbations can be written: 
\be 
\cZ = \int
[d\Phi]\, e^{\ts-S} = \int [d\Phi]\, e^{\ts -\nu \tr\left(e^{\Phi}
-\Phi + \sum_{k=1}^\infty \tbar_{ k}\, \left(e^{ \Phi}e^{
-iX}\right)^{ k}\right)} 
\ee 
with 
\be 
t_{ k} = {1\over\nu k}\tr (e^{-ikX}) 
\label{xkontsmiwa}
\ee 
To understand this better, let us consider the simplest case of $N=1$,
i.e. $1\times 1$ matrices. In this case, the perturbing term becomes:
\be 
\sum_{k=1}^\infty \tbar_{ k}\, \left(e^{\Phi}e^{-iX}\right)^{
k}\to \sum_{k=1}^\infty \tbar_{ k}\, e^{k(\phi -ix)} 
\ee 
The RHS looks like half of the mode expansion of a 2D Euclidean scalar
field. This suggests that we identify $x$ with the (compact) Euclidean
time direction. Indeed, the periodicity is correct, since $\exp(ikx)$
is single-valued at $R=1$ (i.e. $x\to x+ 2\pi$) precisely for integer
$k$. It is plausible that this is how a single D-instanton couples to
closed-string tachyons of negative momentum.

For $N$ D-instantons we need to go back to the noncommuting matrices,
i.e. make the replacement:
\be
e^{k(\phi -ix)}\to \tr \left(e^{\Phi}e^{-iX}\right)^{ k}
\ee
This amounts to a  specific prediction  for the  matrix ordering  on $N$
D-instantons of $c=1,R=1$ string theory. When the exponentials on the
RHS are expanded, we find an infinite sequence of commutator terms
between $\Phi$ and $X$.

The most puzzling aspect of this framework is that $X$ is not a
dynamical variable unlike $\Phi$ (i.e. a random matrix to be
integrated over), but rather a fixed background.  It seems to be
determined self-consistently, like a condensate, in terms of the
positive momentum tachyon couplings, via the Kontsevich-Miwa 
relation Eq.(\ref{xkontsmiwa}).

We can get an indication how this might come about, by going back to
the simple case of $N=1$ and making the following plausible
ansatz for the tachyon couplings:
\be
\Delta S = \sum_{k=1}^\infty \Big(t_{ k}\, e^{k(\phi + ix)} +
\tbar_{ k}\, e^{k(\phi -ix)}\Big)
\label{tachpert}
\ee
Next, assume that $x$ is determined consistently by making the
above term stationary:
\be
{\del\over \del x}\,\Delta S =0
\ee
which leads to:
\be
t_{k}\, e^{k\phi}e^{ikx} = \tbar_{k}\, e^{k\phi}e^{-ikx}
\ee
or
\be
{t_{ k}\over \tbar_{ k}} = e^{\ts -2ikx}
\ee
Now $x\to -x$ is a symmetry of the theory that interchanges $t_1$ with
$\tbar_1$. The unique solution of the above equation that respects
this symmetry is:
\be
t_1 = e^{-ix},\quad \tbar_1 = e^{ix}
\ee
The first of these  equations is a miniature version of the
Kontsevich-Miwa transform,  which is the best we can expect since  we
chose $N=1$. 

Suppose we use the above solution for $t_1$, and think of it as
determining $x$. If we also leave $\tbar_1$ arbitrary (this
procedure does not explain why we should do that) then we find that:
\be
\Delta S \sim\sum_{k=1}^\infty \tbar_{ k} \, e^{k\phi}\,e^{-ikx}
\ee
and $t_1$ is determined by
\be
t_1 = e^{-ix}
\ee
which is more or less the right story for the tachyon couplings of the
$W_\infty$ model at $N=1$. A more complete analysis, particularly at
general $N$, would hopefully shed some light on the non-dynamical
nature of $X$ in this model.

Interestingly there is another matrix model in the literature, the 
normal matrix model\cite{AleKazKos}, which
is very similar to the Penner model but treats $t,\tbar$ in a
manifestly symmetric way. It is defined in terms of a complex matrix
$Z$ that is constrained to be {\em normal}:
\be
[Z,Z^\dagger]=0
\ee
and the matrix integral is:
\be
\cZ(t,\tbar)_{NMM} = \int [dZ\,dZ^\dagger]\,e^{\ts\tr\Big(-\nu ZZ^\dagger 
+(\nu-M)\log ZZ^\dagger
-\nu \sum_{k=1}^\infty (t_k Z^k + \tbar_k {Z^\dagger}^k)\Big)}
\ee
(A variant of this model exists for any finite radius $R$).  The
relationship of this model to the $W_\infty$ model deserves to be
examined in detail, but we will not be able to do this here. Let us
just observe that if we parametrise:
\be
Z = e^{\Phi}\,e^{iX}
\ee
then the unperturbed part of the action depends only on $\phi$ and is
similar to the unperturbed $W_\infty$ model. 

The normality constraint amounts to:
\be
[\Phi,X]=0
\ee
and the tachyon perturbation of this action is:
\be
\Delta  S = -\nu \sum_{k=1}^\infty (t_k\, e^{ik(\Phi+iX)} + \tbar_k\,
e^{ik(\Phi-iX)}
\ee
We see that in this model, both the $\Phi$ and $X$ directions are
represented by  dynamical matrices, and the tachyon perturbation is
similar to the one proposed above in Eq.(\ref{tachpert}).

Indeed, the normal matrix model is supposed to be equivalent to the
$W_\infty$ model at large $N$, as both are equivalent to $c=1,R=1$ by
construction, though a precise equivalence in terms of matrices has
not been demonstrated. In view of our conjecture that the action of
the $W_\infty$ model is that on $N$ D-instantons with the transverse
coordinate $X$ treated as non-dynamical, it seems natural to speculate
that the NMM describes the same system but with dynamical $X$. It is
plausible that both models become equivalent at large $N$, though
superficially they do not appear to be equivalent at finite $N$.

\section{Conclusions}

We have seen that $c=1, R=1$ is a very special theory. It
is completely solved to all orders in perturbation theory, and its
free energy and correlators are known in a genus expansion.
Its partition function is related to the  topology of moduli
space, while the correlators are related to integrable hierarchies. 

All the above information is elegantly encoded in the $W_\infty$
matrix model. Here we have shown that this model maps to a matrix model
with a Liouville-plus-linear interaction. This is proposed to be the
matrix theory on $N$ D-instantons of the $c=1$ string theory. It is
important to confirm whether this proposal is correct and/or needs
modification.

Extensions of these ideas and relationships from $c=1$ to ${\hat c}=1$
(type 0A, 0B) strings will be interesting to pursue. In the
supersymmetric case, there is naively no self-dual radius since the
T-dual of type 0B theory is type 0A theory. However, there is a subtle
``affine quotient'' theory discovered in
Refs.\cite{DifSalZub,DixGinHar} and studied in
Ref.\cite{DouKleKutMalMarSei}, which is T-dual to itself and therefore
does have a self-dual radius. It is this theory to which the
considerations of this article are most likely to generalise.

Traditionally, topological string theories are tied to the
genus expansion by their very definition, in that the partition
function and amplitudes are defined genus by genus. But if, as is
currently believed, the type 0A and 0B string theories truly have a
nonperturbative completion, then the corresponding topological
formulations might have one as well. If the latter can be summarised
as matrix models, it is reasonable to hope that these super-analogues
of the Penner and $W_\infty$ matrix models are not just generating
functions of a genus expansion, but also contain physically meaningful
nonperturbative terms. This could prove extremely important for the
understanding of nonperturbative string theory.

The above observations could also have a bearing on the many
topological theories related to $c=1,R=1$, such as the topological
black hole and topological conifold.

\section*{Acknowledgements}

I am most grateful to the organisers of the IPM Workshop on String
Theory at Bandar-e-Anzali on the Caspian Sea, Iran, for their
invitation to give these lectures. The gracious and gentle hospitality
of the people of Iran made the visit an unforgettable
experience. Helpful conversations with Allan Adams, Rajesh Gopakumar
and Ashoke Sen are acknowledged.

\end{document}